\documentclass{ws-ijqi}
\usepackage{amsmath}
\newcommand{\ket}[1]{\vert #1 \rangle}
\newcommand{\braket}[2]{\langle #1 \vert #2 \rangle}
\newcommand{\ketbra}[2]{\vert #1 \rangle \! \langle #2 \vert}
\newcommand{\Tr}{\hbox{Tr}}
\def\scc{\hbox{\tiny C}}
\def\sct{{\hbox{\tiny T}}}
\begin{document}
\markboth{G. De Cillis, M. G. A. Paris}
{Quantum limits to estimation of photon deformation}
\catchline{}{}{}{}{}
\title{QUANTUM LIMITS TO ESTIMATION OF PHOTON DEFORMATION}
\author{GIOVANNI DE CILLIS}
\address{Fondazione Sacro Cuore, I-20134 Milano, Italia\\ 
giovannidecillis@sacrocuore.org}
\author{MATTEO G A PARIS}
\address{Dipartimento di Fisica, Universit\'a degli Studi di Milano,
I-20133 Milano, Italia.\\
matteo.paris@fisica.unimi.it}
\maketitle
\begin{history}
\received{1 January 2014}
\accepted{4 March 2014}
{Int. J. Quantum Inf. {\bf 12}, 1461009 (2014)}
\end{history}
\begin{abstract}
We address potential deviations of radiation field from the bosonic
behaviour and employ local quantum estimation theory to evaluate the
ultimate bounds to precision in the estimation of these deviations using
quantum-limited measurements on optical signals.  We consider different
classes of boson deformation and found that intensity measurement on
coherent or thermal states would be suitable for their detection making,
at least in principle, tests of boson deformation feasible with current
quantum optical technology.  On the other hand, we found that the quantum
signal-to-noise ratio (QSNR) is vanishing with the deformation itself
for all the considered classes of deformations and probe signals, thus
making {\em any} estimation procedure of photon deformation inherently 
inefficient. A partial way out is provided by the polynomial dependence 
of the QSNR on the average number of photon, which suggests that, in 
principle, it would be possible to detect deformation by intensity
measurements on high-energy thermal states.
\end{abstract}
\date{\today}
\section{Introduction}
\label{s:intro}
In the canonical quantization of the radiation field in the vacuum,
normal modes are associated to quantum harmonic oscillators with mode 
operators $a$ and $a^{\dag}$, obeying the canonical commutation
relations $[a,a^\dag]=1$ for bosonic operators. This is a consequence of
the spin-statistics theorem, which itself has been recently the subject
of experimental verification using either Bose-Einstein-statistics-forbidden 
two-photon excitation in atomic barium \cite{eng10} or all-optical 
superpositions of quantum operations on thermal light fields 
\cite{zav09}. Other tests has been carried out
for different physical systems, e.g. for mesons using the decay $K_L^0
\rightarrow \pi^+ \pi^-$. This decay is usually interpreted as due to 
CP violations, but it may occur without CP violation assuming a 
deformation of Bose statistics for pions \cite{gre91}.  As a matter of fact, 
different tests focus on different aspects of the bosonic nature of the
radiation field, thus showing different levels of precision and posing 
different bounds to the amount of {\em photon deformation}. On the other 
hand, in view of the fundamental interest of the subject, as well as to 
assess the different
strategies to estimate photon deformation, it would be highly desirable to
derive the ultimate bound to the precision of these kind of tests.
\par
In this paper we address potential deviations of radiation field from 
the bosonic behaviour \cite{McFarlane89,Delgado91,Man'ko93,V.I.
Man'koPhys93,Aniello00} and employ local 
quantum estimation theory \cite{Hel76,Mal9X,BC9X,Bro9X,LQE} 
to obtain the ultimate bounds to precision in the estimation
of these deviations using quantum-limited measurements.
We consider different classes of deformations and look for 
optimal measurements able to reveal deviation from the bosonic
behaviour using different families of signals. In particular, 
we address deformed coherent states \cite{Biedenharn89,Quesne02,Celeghini91}, 
thermal states \cite{Isar04}, and 
superposition {\em cat-like} states \cite{Osland01}.
\par
Our approach will be that of addressing the above classes of deformed 
states as families of states parametrized by a deformation-dependent 
parameter, and to employ tools from local quantum estimation theory 
to evaluate the ultimate bounds to precision in the estimation of this 
parameter by quantum-limited measurements \cite{tes11,bru12}. In particular, we evaluate 
the quantum Fisher information (QFI) and the quantum signal-to-noise 
ratio (QSNR), and show that they are achieved by intensity measurements.
This result indicates that estimation of photon deformation at the
quantum limit is in principle feasible with current quantum 
optical technology. However, the quantum signal-to-noise ratio is scaling 
with powers of the deformation itself for all the considered 
classes of deformations, and thus signals with very large energy are
needed to achieve a suitable level of precision. In other words, basic
laws of quantum mechanics make estimation of photon deformation an 
inherently imprecise procedure.
\par
The paper is structured as follows. In Section \ref{s:defs} we 
introduce the classes of deformations and the deformed states we 
are going to consider throughout the paper, whereas in 
Section \ref{s:lqe} we review  local quantum estimation theory
and introduce the quantum Fisher
information and the quantum signal-to-noise ratio. In Section
\ref{s:ldef} we show that intensity measurements are optimal 
for the estimation of photon deformation and evaluate the quantum 
limits to precision for measurements on different deformed states. 
Section \ref{s:outro} closes the 
paper with some concluding remarks.
\section{Deformed coherent and thermal states}
\label{s:defs}
We address tests of deformation based on quantum limited
measurements performed on coherent states and their superpositions 
(Schr\"odinger cat-like states) as well as on states at thermal
equilibrium. More specifically, we consider two kind of possible deformations 
corresponding to commutation relations $a a^{\dag} - qa^{\dag}a = q^{-N}
\mathbb{I}$ \cite{McFarlane89,Biedenharn89} or $a a^{\dag} - qa^{\dag}a = 
\mathbb{I}$ \cite{Arik76}, which will be referred to as $P$ and $M$ 
deformation respectively \cite{KS94}.
In the following we will write $q=1+\epsilon$ and look for precision
bounds on the estimation of $\epsilon$. For $q\rightarrow 1$ the above 
commutation relations reproduce the usual algebra of the harmonic
oscillator.
\par
$P$ and $M$ deformations of the algebra do not modify Fock
number states $|n\rangle$, which coincide with that of the 
harmonic oscillator. On the other hand, $q$-coherent states for 
the $P$ and $M$ deformed algebras are indeed deformed and their
expression is given by 
\cite{Biedenharn89,Quesne02,Celeghini91}:
\begin{align}
\ket{\alpha_{\epsilon}} = \sum_{n=0}^{\infty}{\psi_n(\epsilon)}\ket{n},
\quad
\psi_n(\epsilon)
=\frac{1}{\sqrt{C_{\epsilon}(|\alpha|^2)}}\frac{\alpha^n}{\sqrt{\Delta_n(\epsilon)}}
\label{eq:cs}
\end{align}
where $C_{\epsilon}(|\alpha|^2)$ is a normalization coefficient, and 
the expressions of $\Delta_n(\epsilon)$ for the two deformations 
are given by
\begin{align}\label{eq:def}
\Delta_n(\epsilon)&= \left(-\frac{1}{\epsilon}\right)^n 
g_n(1+\epsilon,1+\epsilon)
& {\rm M}\; {\rm deformation} 
\notag \\ 
\Delta_n(\epsilon)&=
\frac{(-1)^n}{2}
\frac{(1+\epsilon)^{-\frac12 n(n-1)}}{[\epsilon(\epsilon+2)]^n}
g_n(-1,1+\epsilon) g_n(1+\epsilon,1+\epsilon)
& {\rm P}\; {\rm deformation}  
\end{align}
where
$$ g_n(a,b) = \prod_{k=0}^{n-1} (1-a b^k)\,.  $$
Up to the first nonvanishing order in $\epsilon$ we have
\begin{align}\label{expad}
\Delta_n(\epsilon)&= n! \left[1+\frac14 \epsilon\, n\, (n-1)\right]
&{\rm M}\; {\rm deformation}\,, 
\notag \\ 
\Delta_n(\epsilon)&= n! \left[1+\frac1{36} \epsilon^2 n\, (n-1) (2n+5)\right]
& {\rm P}\; {\rm deformation}\,. 
\end{align}
Physical properties of $q$-deformed coherent states, e.g. the photon
distribution \cite{KS94}, are different from those of coherent states
of the harmonic oscillator and thus photon deformation may be detected
by performing quantum limited measurements on known sources of coherent
states, as those provided by classical currents or lasers. Using 
Eq. (\ref{expad}) we obtain the mean number of photon of deformed
coherent states in terms of that of the undeformed ones (up to the first
nonvanishing order in $\epsilon$)
\begin{align}\label{numd}
N&= |\alpha|^2 - \frac12 \epsilon |\alpha|^4 
&{\rm M}\; {\rm deformation}\,, 
\notag \\ 
N&= |\alpha|^2 - \frac12 \epsilon^2 |\alpha|^2 \left(|\alpha|^2 
+ \frac13 |\alpha|^4\right)
& {\rm P}\; {\rm deformation}\,. 
\end{align}
\par
The same line of reasoning is valid for states at thermal equilibrium, whose
deformed versions are expressed in the Fock basis as follows
\cite{Isar04}      
\begin{align}
\nu_\epsilon = \frac{1}{Z_\epsilon}\sum_{n=0}^{\infty}\nu_n(\epsilon)\,\ketbra{n}{n},
\end{align}
with 
\begin{align}
\nu_n(\epsilon)&=
\exp\left\{
-\frac{\beta}{2}\left[\gamma_{1+n}(\epsilon)+\gamma_n(\epsilon)-1\right]
\right\} \qquad Z_\epsilon =\sum_{n=0}^{\infty} \nu_n(\epsilon)
\,
\end{align} 
where we used natural units and unit frequency, $Z_\epsilon$ 
is the partition function and the coefficients $\gamma_n(\epsilon)$
are given by
\begin{align}\label{expg}
\gamma_n(\epsilon)&= 
\frac{(1+\epsilon)^n-1}{\epsilon} \simeq n + \frac12 n(n-1) \epsilon 
&{\rm M}\; {\rm deformation}\,, 
\notag \\ 
\gamma_n(\epsilon)&= 
\frac{(1+\epsilon)^{1-n}}{\epsilon (2+\epsilon)}
\left[(1+\epsilon)^{2n}-1\right]
\simeq n + \frac16 n(n^2-1) \epsilon^2 
& {\rm P}\; {\rm deformation}\,. 
\end{align}
In the limit of vanishing $\epsilon$ we 
recover the undeformed expression $\nu_n(0)=e^{-\beta n}$, with the 
undeformed mean number of thermal photons given by $ n_{\sct}=(e^\beta-1)^{-1}$.
Up to the first nonvanishing order in $\epsilon$ we have
\begin{align}\label{expant}
\nu_n(\epsilon)&= e^{-\beta n}\, (1-\frac12\, \epsilon\, \beta\,n^2)
& {\rm M}\; {\rm deformation}\,, 
\notag \\ 
\nu_n(\epsilon)&= e^{-\beta n}\, \left[1-\frac1{12}\, \epsilon\,
\beta\,n(1+n)(1+2n)\right]
& {\rm P}\; {\rm deformation}\,. 
\end{align}
Also for perturbed thermal states the average number of photons may be expressed
in terms of the unperturbed ones. The formulas are quite cumbersome and
we report the expression for small and large values of $n_\sct$
\begin{align}
N&\simeq n_{\sct} -\epsilon\, \left( 2n_\sct^2 +\frac32 n_\sct -\frac1{12}\right)
&{\rm M}\; {\rm deformation}\quad n_\sct \gg 1 \,, 
\notag \\ 
&\simeq n_\sct +\frac12 \epsilon\, n_\sct \log n_\sct 
&{\rm M}\; {\rm deformation}\quad n_\sct \ll 1 \,, 
\notag \\ 
N&\simeq n_\sct -\epsilon^2\,n_\sct\, \left(3 n_\sct^2 +\frac92 n_\sct +\frac32\right)
&{\rm P}\; {\rm deformation}\quad n_\sct \gg 1 \,, 
\notag \\ 
&\simeq n_\sct +\frac12 \epsilon^2\, n_\sct \log n_\sct 
&{\rm P}\; {\rm deformation}\quad n_\sct \ll 1 \,, 
\end{align}
\par
Finally, let us consider the $q$-deformed analogue of {\em cat} states,
i.e the following superposition of $q$-deformed coherent states 
\cite{Osland01} 
\begin{align}\label{eq:cat}
\ket{C_\epsilon}=\frac1{\sqrt{W_{\epsilon}(|\alpha|^2)}}\left(\ket{\alpha_{\epsilon}}
+\ket{-\alpha_{\epsilon}}\right),
\end{align}
where the normalization is given by 
$$W_{\epsilon} (|\alpha|^2)= 2\left[ 1+ \frac{C_\epsilon(-|\alpha|^2)}{C_\epsilon
(|\alpha|^2)}\right]\,.$$
The average number of photons of an unperturbed cat state is given by
$n_{\scc}=|\alpha|^2\tanh |\alpha|^2$ i.e. 
$n_{\scc} \simeq |\alpha|^4$ for small $|\alpha|$ and 
$n_{\scc} \simeq |\alpha|^2$ for large $|\alpha|$.
For perturbed cat states the average number of photons may be expressed
in terms of the unperturbed ones. Also in this case 
we report the expression for large and small $n_c$
\begin{align}
N&\simeq n_{\scc} -\frac12 \epsilon\, n_{\scc}^2
&{\rm M}\; {\rm deformation}\quad n_c \gg 1 \,, 
\notag \\ 
&\simeq n_{\scc} -\frac12 \epsilon\, n_{\scc}
&{\rm M}\; {\rm deformation}\quad n_c \ll 1 \,, 
\notag \\ 
N&\simeq n_{\scc} -\frac12 \epsilon^2\, n_{\scc}^2
&{\rm P}\; {\rm deformation}\quad n_c \gg 1 \,, 
\notag \\ 
&\simeq n_{\scc} -\frac12 \epsilon^2\, n_{\scc}
&{\rm P}\; {\rm deformation}\quad n_c \ll 1 \,.
\end{align}
\section{Local quantum estimation theory}\label{s:lqet}
\label{s:lqe}
Several quantities that may be of interest in order to characterize a 
quantum systems, as for example entanglement and purity, are nonlinear 
functions of the density matrix and cannot, even in principle, correspond 
to proper quantum observables. The value of these quantities should
be estimated through indirect measurements and thus their determination 
corresponds to a parameter estimation problem \cite{EE,EEE,EEL,edvd}. 
Local quantum estimation theory provides tools to determine the most 
precise estimator, solving the corresponding optimization problem 
\cite{LQE}.
\par
Given a set of quantum states described by the one-parameter family of 
density operator $\rho_{\epsilon}$, the estimation problem is that of
finding an estimator, that is a map $\hat{\epsilon}=
\hat{\epsilon}(\chi)$ from the set of the outcomes $\chi$ to the space of
parameters. Classically, optimal estimators are those saturating the
Cram\'er-Rao inequality $ \textrm{Var}(\epsilon)\geq [M F(\epsilon)]^{-1}$
which bounds from below the variance $\textrm{Var}(\epsilon)=E
[\hat{\epsilon}^2]-E [\hat{\epsilon}]^2 $ of any unbiased estimator of the
parameter $\epsilon$. $M$ is the number of
measurements and $F(\epsilon)$ is the Fisher Information (FI) 
$$ F(\epsilon)= \int\! dx\, p(x|\epsilon)\left[
\partial_\epsilon \ln p(x|\epsilon) \right ]^2\,,$$
where $p(x|\epsilon)$ is
the conditional probability of obtaining the value $x$ when the
parameter has the value $\epsilon$.  The quantum Cram\'er-Rao bound 
is obtained starting from the Born rule $p(x|\epsilon)=
\Tr[\Pi_x\, \rho_\epsilon]$ where $\{\Pi_x\}$ is the probability
operator-valued measure (POVM) describing the measurement.
Upon introducing the Symmetric Logarithmic Derivative (SLD) 
$L_\epsilon$ as the operator satisfying $2 \partial_\epsilon\varrho_\epsilon=  
L_\epsilon \varrho_\epsilon+ \varrho_\epsilon L_\epsilon$ one proves that 
the FI is upper bounded by the Quantum 
Fisher Information (QFI) 
\begin{align}
F(\epsilon)\leq H(\epsilon)&\equiv\Tr[\rho_\epsilon L_\epsilon^2]
=2 \sum_{nm} \frac{\left|\langle \psi_m| \partial_\epsilon
\rho_\epsilon | \psi_n\rangle\right|^2}{\rho_n+ \rho_m}\:,
\label{HF}
\end{align}
where we exploited the diagonal form of $\rho_\epsilon=\sum_n \rho_n |
\psi_n\rangle\langle\psi_n |$ on its eigenbasis.
In turn, the ultimate limit to precision is given by the quantum 
Cram\'er-Rao bound 
$$\textrm{Var}(\epsilon)\geq [M H(\epsilon)]^{-1}\,.$$
The above inequality may be also expressed in terms of 
the signal-to-noise ratio (SNR) $R_{\epsilon} =
\epsilon^2/\text{Var}(\epsilon)$, which is bounded the the
so-called quantum signal-to-noise ratio (QSNR) $Q_\epsilon$
\begin{align}
R_{\epsilon}= Q_{{\epsilon}} = \epsilon^2\,H(\epsilon)\,.\end{align}
The parameter $\epsilon$ is effectively estimable when the corresponding
$Q_{\epsilon}$ is large. In order to obtain a $3\sigma$ confidence interval after
M measurements, the relative error $\delta^2$ has to be
$$\delta^2=\frac{9\text{Var}(\epsilon)}{M\epsilon^2}=\frac{9}{MQ_{\epsilon}}
=\frac{9}{M\epsilon^2H(\epsilon)}\,.$$
Therefore, the number of measurements M needed to achieve a $99.9\%$
($3\sigma$) confidence interval with a relative error $\delta$ scales as
$M_{\delta}=9\delta^{-2} Q_{\epsilon}^{-1}$ \cite{EE}. This means that a vanishing
$Q_{\epsilon}$ implies a diverging number of measurements to achieve a
given relative error, whereas a finite value allows estimation with
arbitrary precision at finite number of measurements.
\section{Quantum limits to estimation of photon deformation}
\label{s:ldef}
We first prove that measuring the intensity of the field is an optimal
detection scheme to estimate the photon deformation on all the classes 
of states we
are considering. This is basically due to
the fact that Fock number states are not affected by deformation. In
order to prove this explicitly let us start from the case of pure
states, 
$$\ket{\psi_{\epsilon}} = \sum_{n=0}^{\infty}\psi_n(\epsilon)\ket{n}\,,$$
for which one has
\begin{align}
F(\epsilon) = 4\sum_{n=0}^{\infty}\frac{(\partial_{\epsilon}|
\psi_n(\epsilon)|^2)^2}{|\psi_n(\epsilon)|^2} = 
4\sum_{n=0}^{\infty}\left[\partial_{\epsilon}\psi_n(\epsilon)\right]^2 
\equiv H(\epsilon)\,.
\end{align}
The first expression is the classical Fisher information for intensity
measurements, while the second one is obtained by specializing Eq.
(\ref{HF}) to pure states.
\par
For thermal states, and more generally for mixed states that are
diagonal in the Fock bases
$\rho_{\epsilon}=\sum_{n=0}^{\infty}\rho_n(\epsilon)\,\ketbra{n}{n}$, 
the quantum Fisher information may be written as
\begin{align}\label{eq:H_dev}
H(\epsilon) = F(\epsilon) + 2\sum_{k\neq h}
\sigma_{kh}|\braket{h}{\partial_{\epsilon}k}|,
\quad \sigma_{hk}=
\frac{\left[\rho_h(\epsilon)-\rho_k(\epsilon)\right]^2}{\rho_h(\epsilon)+\rho_k(\epsilon)}
\,.\end{align}
However, the second term in (\ref{eq:H_dev}) vanishes since
$\ket{k}$ does not depend on the parameter $\epsilon$, and thus
$$H(\epsilon) = F(\epsilon) = 
\sum_{k=0}^\infty\frac{[\partial_{\epsilon}\rho_k({\epsilon})]^2}{\rho_k({\epsilon})}\,.$$
These results are direct consequences of the linear nature 
of $M$ and $P$ deformation, which are not affecting the Fock basis 
\cite{McFarlane89,Delgado91,Celeghini91}. 
\par 
Using the above formulas, we have evaluated the QSNR for the estimation 
of photon deformation by intensity measurements performed on 
different classes of $q$-deformed states. In particular, we have 
addressed coherent, thermal and superposition states in the regime of small 
perturbations $\epsilon \ll 1$ and large energy $N\gg1$, where $N$ is 
the average number of (deformed) photons of the state under investigation. 
In Table \ref{t:Q} we report the behaviour 
of $Q_\epsilon$ (leading order) for different classes of states and for
the two linear deformations introduced above. Owing to the approximations
used for their derivations the formulas are valid for 
$N\epsilon \lesssim 1$.
\begin{table}[h]
\tbl{The quantum signal-to-noise ratio $Q_\epsilon$
(leading order) for the estimation of linear $P$ and $M$ 
photon deformations 
by intensity measurements on different classes of states.\label{t:Q}
}
{\begin{tabular}{@{}cc|c|c@{}}
\toprule
              & coherent & superposition & thermal\\
\colrule
 & & & \\
P deformation 
&$Q^{P}_{\epsilon}\simeq \frac29\epsilon^4N^4$
&$Q^{P}_{\epsilon}\simeq \frac29\epsilon^4N^4$
&$Q^{P}_{\epsilon}\simeq 40\epsilon^4N^4$ \\ 
 & & & \\
\colrule
 & & & \\
M deformation 
&$ Q^{M}_{\epsilon}\simeq \frac18\epsilon^2N^2$
&$ Q^{M}_{\epsilon}\simeq \frac18\epsilon^2N^2$
&$Q^{M}_{\epsilon}\simeq \epsilon^2N^2$\\
 & & & \\
\botrule
\end{tabular}
}
\end{table}
\par
As it is apparent from Table \ref{t:Q} the QSNR for estimation of 
$M$ deformation shows a better scaling than the corresponding quantity
for $P$ deformation, and therefore {\em any} estimation procedure for
$M$ deformation would be more effective than for $P$ deformation.
We also see that the scaling of the QSNR is the same, at
least at the leading order, for all the considered class
of states. In turn, there are no advantages
in using superpositions of coherent states rather than coherent
states themselves. Finally, thermal states offer better performances 
than coherent states due to the larger constant
multiplying the leading order for both $M$ and $P$ deformations.
\par
Our results indicate that the estimation of photon deformation 
is an inherently inefficient procedure, since the QSNR vanishes 
with vanishing parameter $\epsilon$.  On the other hand, the 
polynomial dependence of $Q_\epsilon$ on the average
number of photon suggests that, in principle, it would be possible to
retrieve information about the deformation exploiting 
a suitable amount of energy in the simple measurement of 
the intensity of light on thermal states.
This procedure, however, is only a partial way out since the QSNR
$Q_\epsilon$ is a function of $\epsilon N$ and the formulas in Table
\ref{t:Q} are valid for $\epsilon N \lesssim 1$.
\section{Conclusions}
\label{s:outro}
In conclusion, we have addressed potential deviations of radiation field
from the bosonic behaviour, and used local quantum estimation theory to
obtain the ultimate bounds to precision in the estimation of these
deviations using quantum-limited measurements on optical signals.  We
have considered two examples of linear boson deformation and have shown
that, due to invariance of Fock number under pertubation,
intensity measurements on coherent or thermal states are suitable for
their detection. This result makes, at least in principle, tests of
boson deformation feasible with current quantum optical technology. On
the other hand, we found that the quantum signal-to-noise ratio is
vanishing with the deformation itself (for all the considered classes of
deformation and probe signals), thus making the estimation of photon
deformation an inherently inefficient procedure. The polynomial
dependence of the QSNR on the average number of photon suggests that, in
principle, it would be possible to retrieve information about the
deformation exploiting a suitable amount of energy in the simple
measurement of the intensity of light on thermal states.
\section*{Acknowledgments}
This work has been supported by MIUR (FIRB LiCHIS-RBFR10YQ3H).
MGAP thanks Marco Genovese for discussions.

\end{document}